\documentclass[aip,jap,preprint]{revtex4-1}  
\usepackage{graphicx}  
\usepackage{dcolumn}   
\usepackage{bm}        
\usepackage{subfigure}
\begin{document}


\title{Two-dimensional phononic thermal conductance in thin membranes in the Casimir limit}

\author{I. J. Maasilta}
 \affiliation{Nanoscience Center, Department of Physics, P. O. Box 35, FIN-40014 University of Jyv\"askyl\"a, Finland}
 \email{maasilta@jyu.fi }


\begin{abstract}
We discuss computational analysis of phononic thermal conduction in the suspended membrane geometry, in the experimentally commonly appearing case where heat can flow out radially in two dimensions from a central source. As we are mostly interested in the low-temperature behavior where bulk scattering of phonons becomes irrelevant, we study  the limit where all phonon scattering takes place at the membrane surfaces. Moreover, we limit the discussion here to the case where this surface scattering is fully diffusive, the so called Casimir limit. Our analysis shows that in the two-dimensional case, no analytic results are available, in contrast to the well known 1D Casimir limit. Numerical solutions are presented for the temperature profiles in the membrane radial direction, for several different membrane thicknesses. Our results can be applied, for example, in the design of membrane-supported bolometric radiation detectors. 
\end{abstract}
\maketitle

\section{Introduction}

Thermal conduction in insulators takes place via phonon transport, which is typically limited in bulk by various scattering mechanisms, with phonon-phonon scattering being dominant in the room temperature region \cite{bermanbook}. In this standard case, the thermal conductance $G$, defined as the coefficient between conducted power $P$ and temperature difference, $P=G\Delta T$, can be written in terms of the local thermal conductivity $\kappa$ and geometry dependent terms. However, if one goes to temperatures much below the Debye temperature $T < \theta_D/30$, the phonon-phonon scattering dies away \cite{bermanbook}, and in pure enough samples one is left with just the scattering at the boundary of the sample. In that case, therefore, there is no true local thermal conductivity left. Nevertheless, data is still often plotted in those units, and one has to remember that the thermal conductivity actually depends on the sample dimensions in that case. 

The boundary scattering limit was first analyzed for the case where heat is flowing along a one-dimensional rod log time ago by Casimir \cite{casimir}. In his analysis, the assumption was made that the scattering was fully random at the surface, in other words a phonon impinging on the surface scatters diffusively to any direction in the half-space with equal probability. He obtained the simple and intuitive result that for an infinitely long cylindirical rod, the thermal conductance can be written in terms of the same formula as for bulk scattering, $\kappa=\frac{1}{3} C \overline{v} l$, where $C$ is the specific heat capacity of the phonon gas, $\overline{v}$ is a properly defined average speed of sound \cite{zimanbook}, but with the mean free path $l$ simply given by the diameter of the rod. This diffusive boundary scattering case was later dubbed the Casimir limit, to differentiate it from more general cases, where the boundary scattering is partially or fully specular \cite{bermanziman,zimanbook}. 

In many experimental cases the heat flow is of course one-dimensional, and the standard Casimir results can be compared for example to results on phononic thermal conduction in one-dimensional nanowires at low temperatures \cite{panu,fon2,li,bourg1,bourg2}. However, in some cases the experimental realization is two-dimensional for the heat flow. By two-dimensional we here mean that there is a central heat source, and heat can spread uniformly in all radial directions. This is particularly relevant for the case of membrane-supported low-temperature bolometric radiation detectors \cite{enss}, where the phononic thermal conductance of a suspended membrane is the limiting mechanism for heat dissipation. Recent experiments on suspended SiN membranes \cite{jenni} have clearly demonstrated that the sub-Kelvin thermal conductance in the membrane geometry is not limited by bulk scattering, by studying the temperature profile in the membrane directly. This also agrees with earlier results where different samples were compared \cite{hoevers}, or where the absolute value of thermal conductance was seen to approach the ballistic limit (fully specular surface scattering) \cite{holmes}. For full understanding of the 2D  experiments, one certainly should also understand the Casimir limit for 2D heat flow. Somewhat surprisingly, we did not find any discussion of it in literature.        

In this paper, we thus answer the question how the 2D heat flow behaves in the Casimir limit from a theoretical point of view. We present only a numerical solution: the equation that we derive does not allow an analytical solution, even in the simplest, most symmetric geometry. It remains to be seen whether an analytical solution exists, but based on the discussion here it seems unlikely. There is therefore a major difference between the 2D and 1D Casimir results, and most importantly, {\it one cannot use the bulk formulas for thermal conductivity at all in the 2D case}. In other words, there is no equivalent general rule that the mean free path would be proportional to some dimension of the sample. 
 
\section{Thermal conductance limited by bulk scattering in 2D}

Before we present the equations for 2D Casimir-limited phonon transport, let us discuss the opposite case of 2D heat flow limited by bulk scattering \cite{vu,leivo}. In 1D everyone is familiar with how the extensive variable thermal conductance $G$ depends on the intensive thermal conductivity $\kappa$:  $G=\kappa A/L$, where $A$ is the cross sectional area and $L$ is the length of the sample. What about in 2D? In that case heat flows out radially, and the cross-sectional area is a function of the radial coordinate. Clearly, the equation between $G$ and $\kappa$ is more complicated.  We can derive it by using the Fourier law for the heat flux ${\bf j}=-\nabla T$, and since we assume cylindrical symmetry, only the radial derivative in the gradient of the temperature remains, so that the power is given by

\begin{equation}
P=-A(r)\kappa\frac{dT}{dr}.
\end{equation}

By noting that $A(r)=2\pi rd$, where $d$ is the membrane thickness, separating the variables $r$ and $T$, integrating, and assuming small temperature difference $\Delta T << T$, one can write $P$ as a function of $\Delta T$, and thus identify $G$:

\begin{equation}
G=\frac{2\pi d}{\ln(r_{1}/r_{0})}\kappa,
\label{G2d}
\end{equation}     

where $r_{1} > r_{0}$ are the two radii where the two temperatures in $\Delta T$ are defined. One sees that there is an explicit dependence on the two coordinates left: $G$ is not a function of just the difference of $r_1$ and $r_2$ like in the 1D case. If, furthermore, we assume that we are far from the origin $r_{1}-r_{0} << r_{0}$, we can Taylor expand the logarithm  and simplify Eq. \ref{G2d} to get 

\begin{equation}
G=\frac{2\pi dr_{0}}{L}\kappa,
\end{equation} 

where $L=r_{1}-r_{0}$. This is finally equivalent with the 1D case, as $A=2\pi r_{0} d$. We thus see what we intuitively expect: for small $L$ far from origin, the curvature difference between the two circles is small so that the 2D case approaches the 1D case. However, in many experimental situations this limit is not reached and the full formula (Eq. \ref{G2d}) must be used to determine $\kappa$ from the measured $G$. Moreover, when the temperature difference is not small, one cannot define a linear thermal conductance in terms of the temperature difference anymore. In that case, it is more useful to calculate the power $P$, which is an explicit function of both the local temperature and the bath temperature $T_b$. One can still define a differential thermal conductance $G_{d}=dP/dT$ which agrees with the result in Eq. \ref{G2d}. If one assumes some functional form for the temperature dependence of the thermal conductivity, for example $\kappa=\alpha T^m$, one can invert the power vs. $T$ relation, arriving \cite{jenni} at a simple result for the temperature profile in the bulk scattering case:

\begin{equation}
T(r)=\left [\frac{(m+1)P}{2\pi \alpha d}\ln \left ( \frac{r_{1}}{r} \right )+T(r_{1})^{m+1} \right ]^{1/(m+1)} ,
\label{bulkdiff}
\end{equation} 

where $T(r_{1})$ is a know temperature at radius $r_{1}$, for example at the membrane edge where it would equal the bath temperature.

\section{The approach to study the Casimir limit in 2D}  

As in Casimir's original work \cite{casimir}, we also take the same approach to model a fully diffusive surface. When the scattering on the surfaces is fully diffusive, the phonon emission from the surface element follows the same laws as photon blackbody radiation, but with a redefined phononic Stefan-Bolzmann constant $\sigma$. For an isotropic material, $\sigma$ is given \cite{wolfe} by 

\begin{equation}
\sigma=\frac{\pi^2k_B^4}{120\hbar^3}\left (\frac{2}{c_t^2}+\frac{1}{c_l^2} \right ),
\end{equation}  

where $c_t$ and $c_l$ are the transverse and longitudinal speeds of sound of the material, respectively. These are the only material parameters in the problem. From the analogy with electromagnetic radiation \cite{howell}, we therefore know that the total hemispherically emitted power of a surface element is $\sigma T^4$, and that the emitted intensity from a surface element $dS$ into solid angle $d\Omega$ is given by Lambert's law: $dI=\frac{\sigma}{\pi} T^4 \cos\theta dS d\Omega$, where $\theta$ is the angle that the directional vector makes with the surface normal. In steady state, there must exist a radiative balance between the emitted and absorbed power for each surface element, leading to an equation $\frac{\sigma}{\pi} T_i^4 dS_i \sum_j \cos\theta_i  d\Omega_j = \frac{\sigma}{\pi} \sum_j T_j^4 \cos\theta_j dS_j d\Omega_i+qdS_i$, where the left side is the total emitted power of element $dS_i$ into surface elements $dS_j$, the first term on the right side is the total absorbed power at element $dS_i$ from all other surface elements $dS_j$, and the last term is an external power absorption due to a metallic heater, for example ($q$ has units of W/m$^2$). The solid angle terms can be written in terms of the two surface elements and are thus $d\Omega_j=dS_j\cos\theta_j/R_{ij}^2$ and $d\Omega_i=dS_i\cos\theta_i/R_{ij}^2$, where $R_{ij}$ is the distance between the two surface elements. 

As we are interested in the 2D case with cylindrical symmetry, we can write the above equation in cylindrical coordinates and go to the continuum limit by converting the sums to integrals. At this point, we should also explicitly take into account of the membrane geometry. When considering a surface element on the membrane, we can see that there are two different contributions to the $dS_j$ elements: one is the opposing, parallel membrane surface (no power is exchanged between the surface elements on the same side due to Lambert's law), and the other is the membrane edge at radial coordinate $R$, which has a different contribution because the edge surface is perpendicular to the membrane surfaces. To maintain the cylindrical symmetry, we have to assume a circular membrane edge and a circular heater geometry. Moreover, we simplify the discussion slightly by also assuming that the heater is symmetric with respect to the direction perpendicular to the membrane (i.e both top and bottom surfaces are heated): in this case the upper and lower membrane surfaces have identical temperatures everywhere. This assumption could actually be relaxed by writing the radiative balance equations for the top and bottom surfaces separately, which would lead to a coupled system of equations, which at this point is an unnecessary complication. We also take the temperature of the membrane edge to be fixed by the heat bath, as it is in direct contact with the bulk of the sample. This is a good approximation when the membrane thickness $d$ is much larger than the dominant wavelengths of the thermal phonons involved; for very thin membranes there should be a corrections due to diffractive processes. 

Using the above assumptions, we arrive at an integral equation desribing the radiative balance for 2D Casimir limit as a function of the radial coordinate $r_1$ :

\begin{eqnarray}
 & & T^4(r_1) \left ( \int_{0}^{2\pi}\!\!d\phi_2\int_{0}^{R}\!\!dr_2 r_2\frac{\sin^2\theta_a}{R_{a}^2}+\int_{0}^{2\pi}\!\!d\phi_2\int_{0}^{d}dz \frac{\sin\theta_b(R^2-r_1R\cos\phi_2)}{R_{b}^3} \right ) \nonumber \\ &=&  \int_{0}^{2\pi}\!\!d\phi_2\int_{0}^{R}\!\!dr_2 T^4(r_2) r_2\frac{\sin^2\theta_a}{R_{a}^2} + T^4(R) \int_{0}^{2\pi}\!\!d\phi_2\int_{0}^{d}\!\!dz \frac{\sin\theta_b(R^2-r_1R\cos\phi_2)}{R_{b}^3}+\frac{\pi}{\sigma}q f(r_{1}),
\label{eqfull}
\end{eqnarray}

where the first term on the left is the emitted power from one parallel membrane surface to another and the second from the membrane surface to the membrane edge at $r_1=R$. On the right we have the absorption terms, the first from the other parallel surface, the second from the membrane edge and the third the direct power input, which now has a radial profile $f(r_{1})$. Here $\theta_i$ are the angles between the membrane surface and the distance vectors (with lengths $R_{a}$ and $R_{b}$) in the two cases. From geometry we can easily derive that 
$R_{a}^2=r_1^2+r_2^2-2r_{1}r_{2}\cos(\phi_{1}-\phi_{2})+d^2$, $\sin\theta_{a}=d/R_{a}$, $R_{b}^2=r_1^2+R^2-2r_{1}R\cos(\phi_{1}-\phi_{2})+z^2$, and $\sin\theta_{b}=z/R_{b}$. Thus, it becomes a matter of elementary integration to simplify Eq. \ref{eqfull}, resulting in the following final linear integral equation for the temperature profile $Z(r_{1})=T^4(r_{1})$:

\begin{equation}
Z(r_{1})=\int_0^R\!\!dr_{2}G(r_{1},r_{2})Z(r_{2})+Z(R)H(r_{1})+Cf(r_{1}),
\label{eqshort}
\end{equation}

where the kernel $G(r_{1},r_{2})$ is given by

\begin{equation}
G(r_{1},r_{2})= \frac{2d^2r_{2}(r_{1}^2+r_{2}^2+d^2)}{\left [ (r_{1}^2+r_{2}^2+d^2)^2-4r_{1}^2r_{2}^2  \right ]^{3/2}},
\label{kernel}
\end{equation}      

the "edge" function $H(r_{1})$ as

\begin{equation}
H(r_{1})= \frac{1}{2} \left ( \frac{r_{1}^2+d^2-R^2}{\sqrt{(r_{1}^2+R^2+d^2)^2-4r_{1}^2R^2}} +1 \right ),
\end{equation}      

and the constant $C$ is the normalized external power input $C=2q/\sigma$ so that the function $f(r_{1})$ is one where power is applied and zero where not. 

\section{Numerical results}

We choose to solve equation \ref{eqshort} numerically, as no obvious analytic solution exists. Using the terminology of mathematical literature, equation \ref{eqshort} can be classified as a linear Fredholm equation of the second kind \cite{numrec}. For the numerics it is very important how the kernel behaves. As is seen from Eq. \ref{kernel}, it is not symmetric, nor is it a function of just a difference of the variables. It is, however, non-singular at $r_{1}=r_{2}$, so that standard numerical techniques for these type of integral equations work. We have chosen to use the so called Nystr\"om method \cite{numrec}, which uses the Gauss-Legendre quadrature rule for discretization of the integral, and trianglular decomposition techniques for the inversion of the obtained linear equations. By trial and error we have seen that the details of the problem influence how many points are needed in the discretization. The higher the aspect ratio between membrane radius and thickness, the higher number of points must be used. Here, we we had to go up to 900 points  (900 by 900 matrix inversion) to get accurate results for the thinnest membranes studied here (aspect ratio $R/d=250$).  We note that thinner membranes would require even more points. 

\begin{figure}
\includegraphics[width=\textwidth]{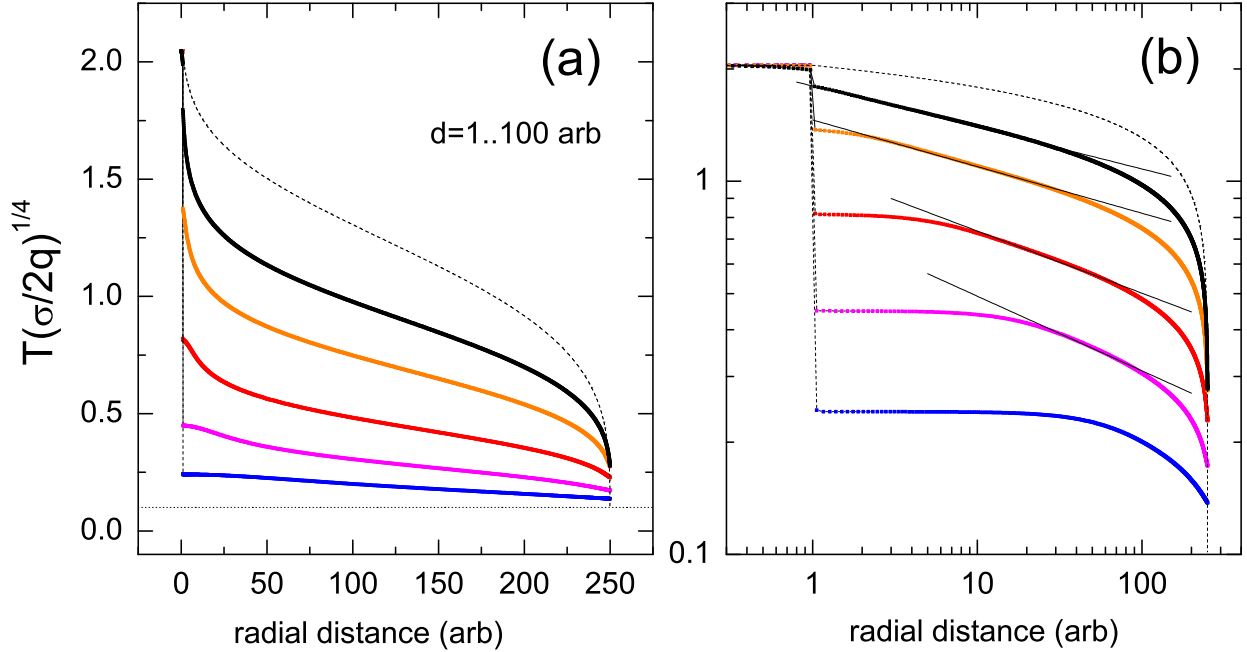}
\caption{[Color online] Calculated temperature profiles for 2D Casimir heat conduction (points) for $R=250$ and $d=1,3.16,10,31.6,100$, where $R$ and $d$ have arbitrary units in (a) linear scale, (b) log-log scale.  Higher curves have lower $d$. Dashed line shows the bulk diffusive result, Eq. \ref{bulkdiff}, and dotted line the value of bath temperature used. In (b), the solid lines are power law fits with exponents -1/9, -1/8, -1/6, -1/5 from top to bottom.} \label{fig1}
\end{figure}

In Fig. \ref{fig1} we present the results for the temperature profiles in the radial direction for several different aspect ratios ranging from $R/d=2.5$  to $R/d=250$, keeping $R$ constant, in linear (a) and log-log scales (b), with a heater radius $r_{h}=1$. As there is no intrinsic length scale in the problem, all results can be plotted in arbitrary units; in other words the plots could represent results for $R=250 \mu$m and $d$ ranging from 1 $\mu$m to 100 $\mu$m, or any other length scale. Temperature is also plotted in the naturally scaled units, which depend on the material through $\sigma$ and heater power through $q$, and heater powers are selected in  such a way that the temperature at the center $r=0$ is the same for all membrane thicknesses. In addition to the numerical Casimir-limit results, we also plot the bulk diffusive case of Eq. \ref{bulkdiff}. 

First, we see that the bulk diffusive result gives higher temperatures than all the calculated Casimir cases (except very near the mebrane edge at $r=250$), and that the Casimir results clearly produce different shapes for the profiles. The profiles also clearly depend on the aspect ratio, so that there is no size-independent universal shape for the Casimir case. This is in contrast with the bulk diffusive and ballistic limits \cite{jenni}. The second very obvious feature is that the temperature profile is not necessarily continuous, and there are jumps both at the heater edge at $r=1$ and at the membrane edge at $r=250$. This is in agreement with Monte Carlo simulations of the 1-D Casimir limit \cite{klitsner}, where temperature jumps at the rod ends were also seen. The temperature jump at the heater edge grows significantly in size when the membrane becomes thicker, so much so that in the thickest membranes studied here, a whopping 93 \% of the temperature drop in the sample occurs right at the heater edge. The drop at the membrane edge is much more modest in all the cases, now being largest ( 9 \%) for the thinnest membrane. The growth of the drop at the heater edge is intuitively understandable: for thick membranes the heater can emit to a much larger solid angle, therefore distributing the heat load much more widely and non-locally, resulting in much cooler membrane outside of heater area.

By plotting the profiles in log-log scale, Fig. \ref{fig1} (b), one can also try to infer phenomenological dependecies for the temperature decay as a function of radial distance. The Casimir-limit temperature profiles outside the heater area seem to have several regions, depending on $d$: there is an initial flat region for the thicker membranes (extending up to $r \sim 30$ for $d=100$), then an intermediate power law decay, where the exponent is very weak and depends on $d$, and then finally a stronger decay beyond $r \sim 50$, resembling the bulk diffusive limit. We have fitted simple power laws $r^{-b}$ to some of the intermediate regions of the plots, where $b$ was varying from 1/9 to 1/5. However, no clear power-law region exisits for the thickest membrane result. 

  \begin{figure}
\includegraphics[width=\textwidth]{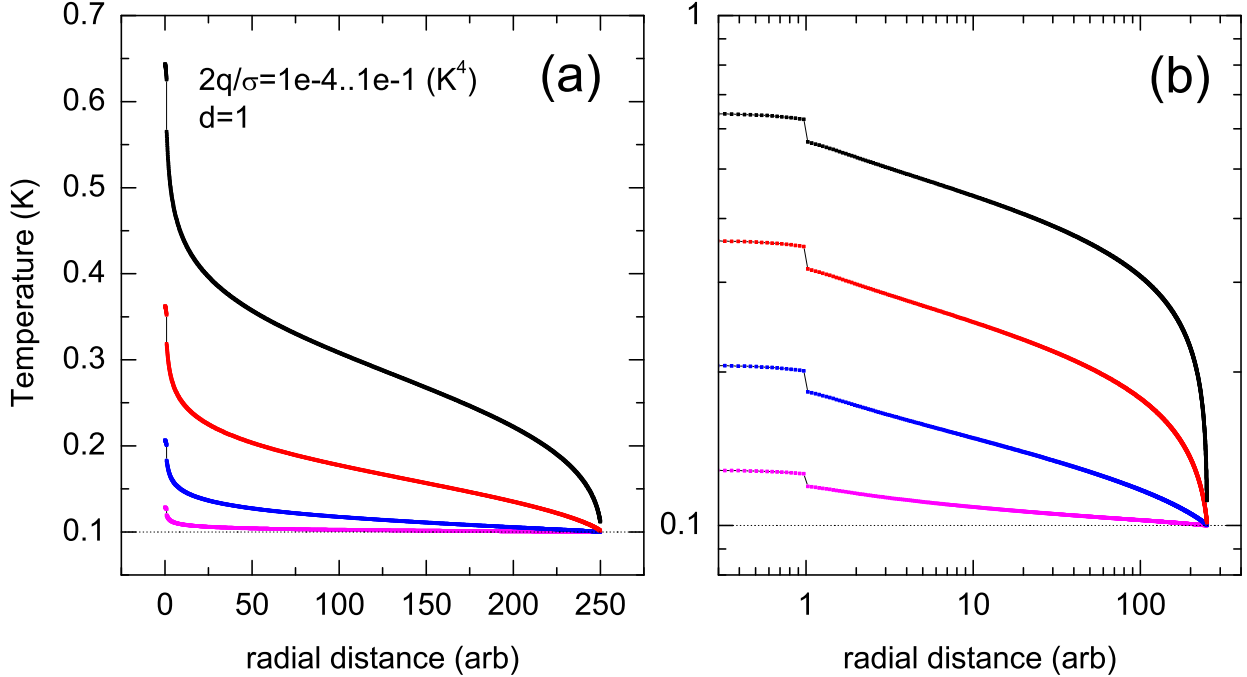}
\caption{[Color online] Calculated temperature profiles for 2D Casimir heat conduction (points) for $R=250$, $d=1$, and $r_h$=1, with varying heat input power density $2q/\sigma=10^{-4}$, $10^{-3}$, 0.01, and 0.1 K$^4$ in (a) linear scale, (b) log-log scale.  Higher curves have higher $q$. The dotted line is the value of bath temperature 0.1 K used. } \label{fig2}
\end{figure}
The profiles presented in Fig. \ref{fig1} are universal in terms of the power input (as implied by the scaling used for $T$) as long as the temperatures at the membrane edge remain higher than the bath temperature $T_{edge}^4 >> T_{bath}^4$.  If that is not the case, the shapes are distorted and become flatter. We show this effect in Fig. \ref{fig2}, where results for $d=1$ with low input power levels are shown. Clearly, with low input power, the temperature profile is very flat and does not show a discontinuous drop at the membrane edge, until a higher power level is reached ($2q/\sigma=0.1$ for this aspect ratio). The power law decay, on the other hand, is not strongly affected until at the very lowest power input levels. The strongest difference between the low power profiles (Fig. \ref{fig2}) and the high power profile (Fig. \ref{fig1}) is in the third region beyond $r=50$, where the stronger, bulk-like temperature decay disappears for low input powers, and becomes linear. Similar behavior takes place for thicker membranes (not shown).                     

  \begin{figure}
\includegraphics[width=\textwidth]{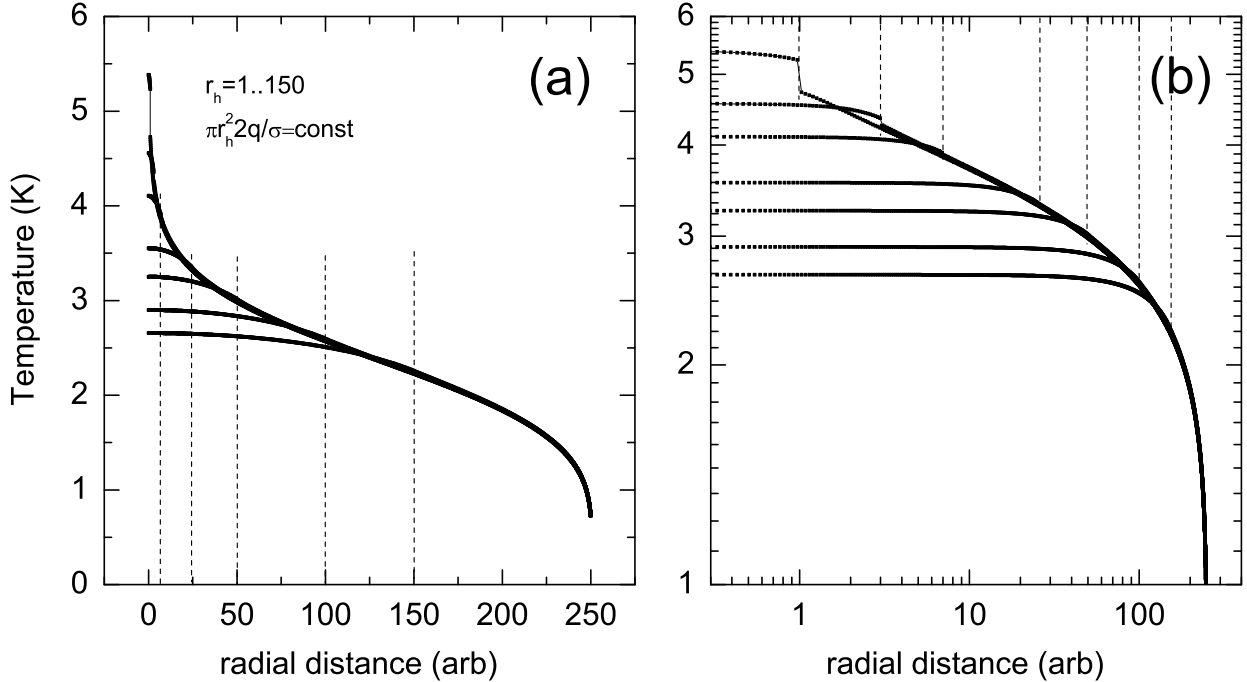}
\caption{[Color online] Calculated temperature profiles for 2D Casimir heat conduction (points) for $R=250$, $d=1$, and constant total heating power $r_{h}^2 2q/\sigma= 490$ K$^4$(arb)$^2$, but with varying heater radius $r_{h}=$1,3,7,25,50,100,150 in (a) linear scale, (b) log-log scale.   The dashed lines mark the values of the heater radii.  } \label{fig3}
\end{figure}

Finally, we want to investigate how the temperature profiles depend on the heater size. Fig. \ref{fig3} shows the calculated temperature profiles for $d=1$, $R=250$, with varying heater size from $r_{h}=1$ to $r_{h}=150$, but by keeping the total power constant. Clearly, all the curves come together outside of the heater region; only the temperature profile within the heater are is affected, with heater temperature naturally rising with rising power density. We also point out that the temperature wihin the heater region is in no way constrained to be constant. In fact, as the larger heater size results clearly show, there is a sizeable temperature gradient wihin the heater area. The geometry only demands that the temperature gradient at the center $r=0$ is zero. Also, the finite temperature drop at the heater edge disappears for larger heater radii $r_{h}>3$.    

\section{Discussion and conclusions}

We studied theoretically two-dimensional phonon transport in suspended membranes in the limit where all the scattering takes place on the membrane surfaces, and with the assumption that all the scattering is diffuse, the so called Casimir limit. What we mean by two-dimensionality in this context is that there is a central heat source, from which heat emanates, spreading out into the membrane and the heat bath, which is located at the edge of the membrane some distance away. We showed that already in the more general case of 2D heat low with bulk scattering, there is no way to describe in general the heat transport properties with just thermal conductivity, as thermal conductance depends explicity on the positions where temperature is measured. We also showed that in the Casimir limit, the situation is complicated even further so that there is no universal temperature profile, but instead it depends on the aspect ratio between the membrane radius and thickness. We presented the derivation of an integral equation for the temperature profile from power balance at the surfaces, and proceeded to solve the equation numerically for several cases. 

 Our formulation assumed a full cylindrical symmetry for simplification. However, in reality suspended membranes tend to have rectangular edge geometries, and also rectangular shaped heater regions. The rectangular case would be thus interesting to study, but it is unfortunately more complex, resulting in fully two-dimensional integral equation. 
 Our results  should, nevertheless, describe fairly accurately even the rectangular geometry, if temperatures are not probed too near the heater or too near the membrane edge. Another future extension of this work could also be the inclusion of specular scattering probability component, as it is known that for many smooth samples a large portion of the scattering on the surfaces is specular \cite{klitsner}. By increasing the specular scattering for thin membranes, one then crosses over to the regime where phonon modes are modified due to interference, and become the so called Lamb modes with non-linear dispersion relations \cite{thomas}. It is known that the Lamb modes affect heat conduction strongly at least in the ballistic limit \cite{thomas2}, therefore the cross-over from Casimir to ballistic conduction should also be studied further.

\section*{Acknowledgements}
 This research was supported by Academy of Finland project number 128532. We thank R. van Leeuwen for many helpful discussions about solving intergral equations and T. K\"uhn for discussions on phonon transport.

\end{document}